\documentclass[conference,letterpaper]{IEEEtran}
\IEEEoverridecommandlockouts
\usepackage{amsmath,amssymb,amsfonts}
\usepackage{algorithmic}
\usepackage{graphicx}
\usepackage{textcomp}
\usepackage[OT1]{fontenc}
\usepackage{pifont}
\usepackage{multirow}
\usepackage{threeparttable}
\usepackage{booktabs}
\newcommand{\tabitem}{~~\llap{\textbullet}~~}
\usepackage[dvipsnames,svgnames]{xcolor, colortbl}
\newcommand{\etal}{{\em et al.\ }}

\definecolor{LightCyan}{rgb}{0.88,1,1}
\def\BibTeX{{\rm B\kern-.05em{\sc i\kern-.025em b}\kern-.08em
    T\kern-.1667em\lower.7ex\hbox{E}\kern-.125emX}}
\begin{document}
\bstctlcite{IEEEexample:BSTcontrol}
\title{Establishing Cyber Resilience in Embedded Systems for Securing Next-Generation Critical Infrastructure\\
}
\author{
\IEEEauthorblockN{
Fahad Siddiqui,
Matthew Hagan,
Sakir Sezer
}
\IEEEauthorblockA{
\textit{The Centre for Secure Information Systems (CSIT), Queen's University Belfast}\\
Belfast, United Kingdom \\
f.siddiqui, m.hagan, s.sezer@qub.ac.uk}
}
\maketitle
\begin{abstract}
 

The mass integration and deployment of intelligent technologies within critical commercial, industrial and public environments have a significant impact on business operations and society as a whole. Though integration of these critical intelligent technologies pose serious embedded security challenges for technology manufacturers which are required to be systematically approached, in-line with international security regulations.

This paper establish security foundation for such intelligent technologies by deriving embedded security requirements to realise the core security functions laid out by international security authorities, and proposing microarchitectural characteristics to establish cyber resilience in embedded systems. To bridge the research gap between embedded and operational security domains, a detailed review of existing embedded security methods, microarchitectures and design practises is presented. The existing embedded security methods have been found ad-hoc, passive and strongly rely on building and maintaining trust. To the best of our knowledge to date, no existing embedded security microarchitecture or defence mechanism provides continuity of data stream or security once trust has broken. This functionality is critical for embedded technologies deployed in critical infrastructure to enhance and maintain security, and to gain evidence of the security breach to effectively evaluate, improve and deploy active response and mitigation strategies. To this end, the paper proposes three microarchitectural characteristics that shall be designed and integrated into embedded architectures to establish, maintain and improve cyber resilience in embedded systems for next-generation critical infrastructure.

\end{abstract}


\section{Introduction}
\label{sec:Introduction}
Proliferation of intelligent connected technologies are opening venues to new service and computing models, providing diverse socio-economic benefits. These technologies are giving rise to wide range of intelligent applications including smart home, smart city, smart grid and intelligent transportation systems~\cite{Wan2012},~\cite{Sharma2019}. Estimates from market leading industry predict that intelligent connected technologies will proliferate to a trillion devices by 2035~\cite{arm2017}. This rapid growth of intelligent consumer and industrial solutions is leading to significant growth in smart embedded devices, such as wearables and critical infrastructure components, that provide information and communication functions to the users and businesses. 

These smart embedded devices will be integrated and deployed in public and private environments for commercial and non-commercial 
purposes, to enhance business and consumer experiences by sharing and analysing generated data~\cite{ncsc2016},~\cite{dhs2016}. This data can be used in a variety of ways, enhancing the customer's experience, bringing new business models and market opportunities using artificial intelligence, machine learning and data analytics, to make better informed decisions. However, where this sharing of data brings benefits and opportunities, it simultaneously presents risks~\cite{Ukil2011}. The large-scale integration and deployment of smart embedded devices and related services within critical infrastructure environments to control critical tasks, poses serious design, supply chain, security and safety challenges~\cite{Sharma2019},~\cite{Ravi2004},~\cite{Apthorpe2017},~\cite{Serpanos2013}.

As reliance on these technologies has grown, opportunities have arisen for adversaries to attack and compromise public and commercial critical infrastructure systems~\cite{ncsc2016},~\cite{dhs2016},~\cite{Ukil2011}. Therefore, 
international government agencies have released cyber security regulations~\cite{NISTCSF},~\cite{NISTRMF},~\cite{NCSCNIS} to curtail this problem by advocating businesses and technology manufacturers to comply and adhere to these regulations.
These cyber security regulations pose a need for smart embedded devices and intelligent technologies to be \textit{Cyber Resilient}. Device manufacturers therefore should design, develop and deploy security within their products to maintain compliance, consumer confidence and market share. However this need for harnessing security to comply with cyber security regulations, has compelled embedded designers and security architects to deploy defences that are often ad-hoc and passive in nature. As they have been designed to mitigate a certain class of known attacks~\cite{Meng2018}. Nevertheless, this strategy has been found vulnerable and compromised due to software vulnerabilities, microarchitectural weaknesses and poor use of secure design practices~\cite{Apthorpe2017},~\cite{Choras2016},~\cite{McClintic2018},~\cite{Chen2017},~\cite{Lipp2017},~\cite{Kiriansky2018}. 

Open literature and reported events show that attack methods are evolving and becoming sophisticated, software vulnerabilities are inevitable, embedded architectures are insecure and are therefore susceptible to diverse attacks~\cite{Meng2018},~\cite{Chen2017}. A successful launch of an attack on a device can expose private and confidential data of the user and enterprise to adversaries. To best of our knowledge, no existing embedded security microarchitecture or defence mechanism provides continuity of data stream and the information that can be extracted to gain and establish an evidence caused by the security breach for \textit{Cyber Forensics}.

Considering these diverse cyber security challenges, there is a need for adopting a holistic rather than continue pursuing passive approach to achieve cyber resilience in embedded systems. The architecture shall harness, maintain and ensure design and operational security. Moreover, it shall be capable of both detection and recovery from a launched attack, and preserve crucial security requirements of embedded device deployed within public and private critical environments.

This paper will present the core security functions set out by the international security authorities in Section~\ref{sec:NIST}. A comprehensive review of existing embedded security practices and mapping of core security functions to existing embedded security landscape will be presented in Section~\ref{sec:sec-req} which will be used to derive the security requirements of a cyber resilient embedded system. The shortcomings of well established embedded security microarchitectures will be discussed in Section~\ref{sec:defenselayer} and microarchitectural characteristics of a cyber resilient embedded system will be proposed in Section~\ref{sec:characteristics}.





\section{Cyber Resilience \& Cyber Security Regulations}
\label{sec:NIST}
Currently, major differences exist in the way companies are using technologies and adopting security practices into their design, development and operational processes
making it more difficult to mitigate and fight against cyber attacks~\cite{Choras2016}. This problem has been elevated by the lack of adoption of security and cyber resilient posture by the stakeholders. 
IT Governance is a global provider of cyber risk and privacy management solutions that defines \textit{Cyber resilience}~\cite{CyberR} as:

\begin{center}
\textit{"The ability of a system to identify, prevent, and respond to cyber attacks, intended to disrupt the system's operational capabilities while maintaining confidentiality and integrity of the data"}
\end{center}

To streamline these security issues, the \textit{National Institute of Standards and Technology} (NIST) and the \textit{National Cyber Security Centre} (NCSC) have released the following frameworks and regulations to improve security:

\begin{itemize}
	\item NIST \textbf{R}isk \textbf{M}anagement \textbf{F}ramework (RMF)
	\item NIST \textbf{C}yber \textbf{S}ecurity \textbf{F}ramework (CSF)
	\item NSCS Security of \textbf{N}etwork and \textbf{I}nformation \textbf{S}ystems Regulations (NIS)
\end{itemize}

\begin{figure}[h]
\centerline{\includegraphics[scale=0.31]{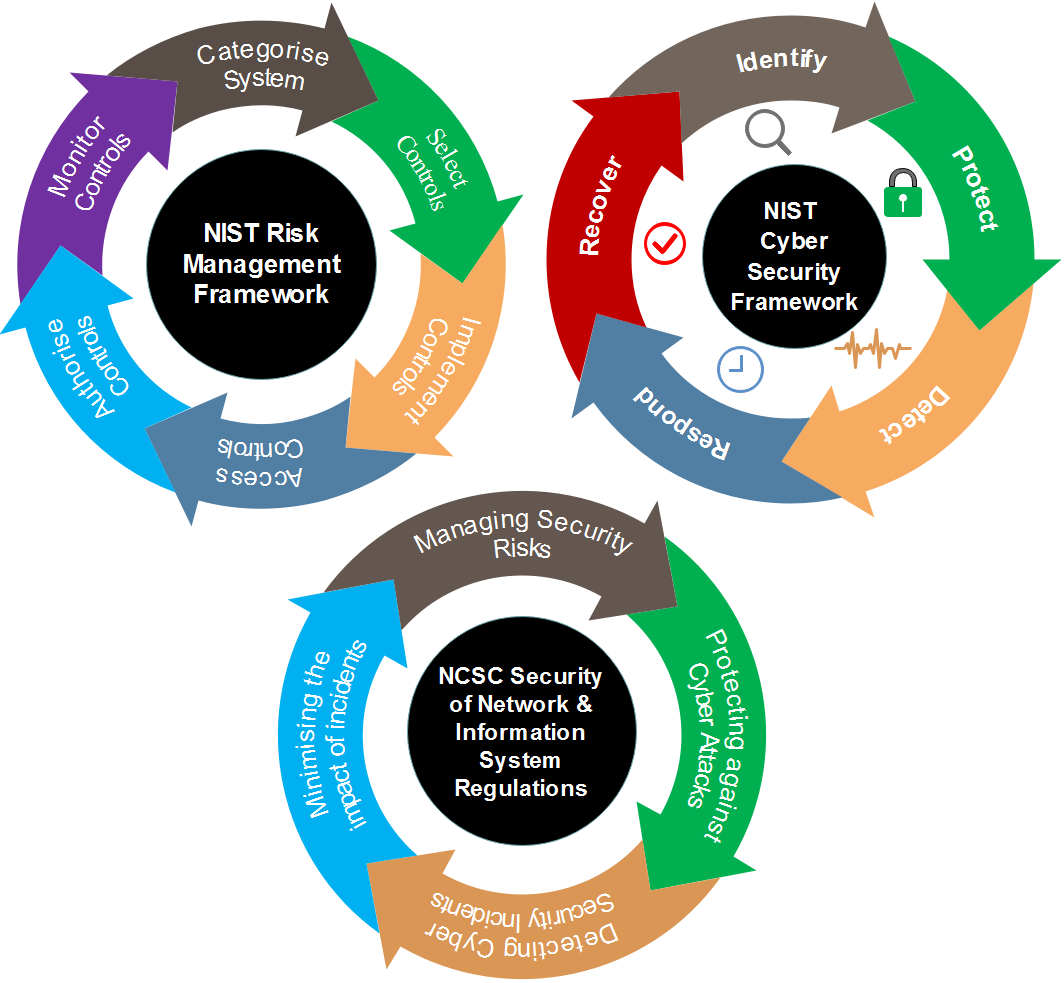}}
\caption{Core security functions, principles and activities of NIST Risk Management Framework~\cite{NISTRMF}, NIST Cyber Security Framework~\cite{NISTCSF} and NCSC Security of Network and Information System Regulations~\cite{NCSCNIS}. 
}
\label{fig:NIST-NCSC}
\end{figure}

The NIST \textit{Risk Management Framework}~\cite{NISTRMF} is a guidance document designed to help organisations and enterprises assess and manage risks to their information and infrastructure. It enables a process that integrates security and risk management activities within the system development life cycle as shown in Figure~\ref{fig:NIST-NCSC}. It provides means to \textit{select}, \textit{implement}, \textit{assess}, \textit{authorise} and \textit{monitor} controls. This involves identification of critical components based on their security requirements followed by selection and implementation of effective monitoring controls, that are aligned with the system's operational behaviour. This process enables security architects to identify risks, select suitable mitigation strategies and deploy countermeasures. This also avoid vulnerabilities which might be overlooked in product functional specification.

The NIST \textit{Cyber Security Framework}~\cite{NISTCSF} 
aims to improve the security of critical infrastructure from cyber attacks. It provides a set of guidelines for technology manufacturers to follow and better prepare to handle cyber attacks, particularly where a lack of security standardisation exists. The framework defines five core security functions (\textit{identify, detect, protect, respond} and \textit{recover}) to establish, maintain and improve cyber resilience as illustrated in Figure~\ref{fig:NIST-NCSC}.

The primary focus of NCSC Security of \textit{Network and Information Systems} (NIS)~\cite{NCSCNIS} regulation is to respond to rising cyber security challenges faced by public/private organisations and enterprises by minimising the risks of disruption to services caused by the failure of digital technologies. One of the key objectives is to establish and improve cyber resilience of intelligent technologies by identifying and managing risks of potential causes of failure by gaining and establishing an evidence of the caused security breach. For this purpose, the regulation introduces four security principles \textit{(managing security risks, protecting against cyber attacks, detecting cyber security incidents and minimising the impact of incidents)} as shown in Figure~\ref{fig:NIST-NCSC}.

The discussed frameworks and regulations advocates that it is essential for technology manufacturers and their involved partners, including semiconductor and original equipment manufacturers (OEMs), to manage their risks by implementing appropriate and proportionate embedded security measures for next-generation critical infrastructure.

\begin{table*}[t]
\centering
\scriptsize
\caption{Association of NIS security principles and CSF core security functions, their respective operational security and derived embedded security requirements for a cyber resilient embedded system. Mapping of existing embedded security landscape on to the driven security requirements is also presented.}
\label{tab:security-requirements}
\begin{tabular}{|l||l||l|l||l|}
\hline
\textbf{NSCS NIS} & \textbf{NIST CSF}  & \textbf{Operational Security} & \textbf{Cyber Resilient} & \textbf{Existing Embedded Security Practices,} \\
\cite{NCSCNIS}	&  \cite{NISTCSF} & \textbf{Requirements} & \textbf{Embedded Security Requirements} &	\textbf{Methods and Microarchitectures}\\
\hline
\hline
\rowcolor{Thistle}	&    & \textbf{Asset Management}	 & \textbf{Embedded Security Modelling}	& \\
\rowcolor{Thistle} \textbf{Managing}	&    & \tabitem Understand and Assess & \tabitem Risk Assessment	& \ding{74} STRIDE, PASTA, CVSS, DREAD, HARA\\
\rowcolor{Thistle} \textbf{Security} & \textbf{Identify} & \tabitem Identify Risks	& \tabitem Threat and Security Modelling & \ding{118} IEC 61508, ISO2626 (ASIL A-D), ISO/IEC 15408\\
\rowcolor{Thistle} \textbf{Risks}  &	 & \tabitem Prioritise and Evaluate	& \tabitem Attack surface identification & \ding{118} Common Criteria, FIPS 140-2, ETSI TVRA\\
\rowcolor{Thistle}	& & \tabitem Comply and Review	& \tabitem Secure-by-design practises & \ding{118} ISO/IEC 27005, SAE J3061, ISO/IEC 27001\\
\hline
\hline
\rowcolor{BlanchedAlmond} & & \textbf{Awareness Control} & \textbf{Protection Method}	& \\
\rowcolor{BlanchedAlmond}\textbf{Protecting} & & \tabitem Protect Data & \tabitem Chain of Trust	& Root of Trust, Trusted Technologies, \ding{74} Secure boot\\
\rowcolor{BlanchedAlmond} \textbf{against} & \textbf{Protect}  & \tabitem Protection Technology	& \tabitem Data Confidentiality and Integrity & \ding{74} AES, ECC, RSA, EDSA, ECCDSA, SHA, SSL\\
\rowcolor{BlanchedAlmond} \textbf{Cyber attack} &	 & \tabitem Manage \& Adopt	& \tabitem Secure Provisioning \& Attestation & \ding{74} Digital Certificate, Public-Private Key Infrastructure\\
\rowcolor{BlanchedAlmond} &	 &	& \tabitem Isolation and Segregation & \ding{74} ARM TrustZone, Intel SGX\\
\hline
\hline
\rowcolor{SkyBlue} &	 & \textbf{Event Discovery}	& \textbf{Detection Method} &\\
\rowcolor{SkyBlue} \textbf{Detecting} &	 & \tabitem Discover \& Determine & \tabitem Platform Security Architecture & \ding{74} ARM Platform Security Architecture\\
\rowcolor{SkyBlue} \textbf{Cyber Security} & \textbf{Detect}   & \tabitem Continuous Monitoring 	& \tabitem Trusted Execution Environment & \ding{74} Global Platform, ARM TEE, QSEE, Kinibi\\
\rowcolor{SkyBlue} \textbf{Incidents} &  & \tabitem Detect Anomalies	 & \tabitem Static \& Dynamic Flow Integrity &  \ding{74} Dover~\cite{Dover206}, \ding{89} ARMHEx~\cite{Wahab2017}\\
\rowcolor{SkyBlue} &	 & \tabitem Alert Events	& \tabitem Access Control and Policing & \ding{89} SECA~\cite{Coburn2005}\\
\hline
\hline
\rowcolor{LightGreen} \cellcolor{YellowOrange} &	& \textbf{Response Planning}	 & \textbf{Response Method} &\\
\rowcolor{LightGreen} \cellcolor{YellowOrange}	 &	 & \tabitem Analyse detected events & \tabitem Platform Security Manager & \ding{74} Trusted Platform Module\\
\rowcolor{LightGreen} \cellcolor{YellowOrange} \textbf{Minimising the} & \textbf{Respond}  & \tabitem Response Strategy	 & \tabitem Physical Security & \ding{74} Side-channel countermeasure\\
\rowcolor{LightGreen}  \cellcolor{YellowOrange} \textbf{impact of} &	 & \tabitem Mitigation Strategy	 & \tabitem Passive countermeasure & \ding{74} Reboot, Reset, Key zeroisation\\
\rowcolor{LightGreen} \cellcolor{YellowOrange} \multirow{2}{*}{} &	 & \tabitem Report \& Improve & \tabitem Active countermeasure & \\
\cline{2-5}
\rowcolor{MistyRose} \cellcolor{YellowOrange}  \textbf{cyber security} &	 & \textbf{Recovery Planning}	& \textbf{Recovery Method} & \\
\rowcolor{MistyRose} \cellcolor{YellowOrange} \textbf{incidents} &	 & \tabitem Repair and Update	& \tabitem Roll-back and Roll-forward & \ding{74} Secure Firmware Update, On-the-air update\\
\rowcolor{MistyRose} \cellcolor{YellowOrange} & \textbf{Recover}  & \tabitem Improve and Train	& \tabitem Fault avoidance and tolerance & \ding{74} Single event upset, Parity, Error Correction Codes\\
\rowcolor{MistyRose} \cellcolor{YellowOrange} &	 &   \tabitem Communicate	 & \tabitem Static and Dynamic Redundancy & \ding{74} Hardware/Software redundancy, Process pairs \\
\rowcolor{MistyRose} \cellcolor{YellowOrange} &	 &	 \tabitem Evidence Collection & \tabitem System Monitoring & \ding{74} Voltage, clock and temperature monitors\\
\hline
\end{tabular}
\begin{tablenotes}
\item \ding{118} International Standard ; \ding{74} Commercially Available ; \ding{89} Academic Research Frameworks/Solutions
\end{tablenotes}
\end{table*}

\section{Security Requirements of\\ Cyber Resilient Embedded System}
\label{sec:sec-req}

Cyber Resilience in embedded systems can be achieved by identifying the security requirements and incorporating them into the product life cycle.
Intrinsically, the discussed Cyber Security regulations in Section~\ref{sec:NIST} do not render security requirements for cyber resilient embedded system. Instead they yield a blueprint which can be used to articulate and derive security requirements for cyber resilient embedded system. Table~\ref{tab:security-requirements} shows the association between NIS security principles and CSF core security functions and their operational security requirements. This includes \textit{asset management, awareness control, event discovery, response planning} and \textit{recovery planning} which are used to derive the security requirements of a cyber resilient embedded system. To bridge the research gap between information security and embedded security, the mapping of each driven embedded security requirement onto existing embedded security landscape is presented in Table~\ref{tab:security-requirements}.
\vspace{0.10cm}


\noindent \textbf{\textcircled{1} IDENTIFY} and manage cyber security risks by conducting \textit{asset management} which involves detailed understanding of an application use case and respective deployment scenario. This requires decomposition of system components and evaluation of their interactions with internal and external entities to identify their associated risks and threats~\cite{Khan2017}. This is followed by evaluating and prioritising tasks, where potential damage to the system and its infrastructure for each identified threat. 

In embedded domain this process is well established which involves creating an abstraction of the embedded system~\cite{Ray2018} known as \textit{Threat and Security Modelling}~\cite{Hagan2018},~\cite{Hagan2018DSC}. This builds profiles of a potential attacker, their goals and methods, which then used to define and deploy countermeasures either to mitigate, minimise the impact of the attack or making less attractive for an attacker. Table~\ref{tab:security-requirements} list some of the risk and threat assessment modelling methods and international standards. They provide detailed guidelines and specifications to model, implement and comply diverse security in embedded systems.\vspace{0.10cm}

\noindent \textbf{\textcircled{2} PROTECT} against cyber attacks by introducing system \textit{awareness control}. This required deployment of appropriate data security and protection methods to build a security foundation based on the principles of information assurance~\cite{Ray2018}.
\begin{itemize}
\item \textbf{Confidentiality}: Ensuring that information is disclosed only to intended individuals, entities and processes. 
\item \textbf{Integrity}: Maintaining and assuring the accuracy and completeness of information over its life cycle.
\item \textbf{Availability}: Ensuring that information must be available when needed by individuals, entities and processes.
\item \textbf{Authentication}: Ensuring that information is accessible by only authorised individuals, entities and processes. 

\end{itemize}

In the embedded security domain, well-established cryptography-based protection methods have been published as shown in Table~\ref{tab:security-requirements}. These protection methods require strong trust anchor to establish and maintain confidentiality, integrity and authentication~\cite{Ray2018},~\cite{Maene2018},~\cite{SiddiquiIET2018}. In addition, embedded access control protection methods such as ARM TrustZone and Intel SGX have been widely used to achieve resource isolation and segregation by dividing system into subsystems and isolating their memory spaces.\vspace{0.10cm}

\noindent \textbf{\textcircled{3} DETECT} cyber security incidents using \textit{event discovery} methods. This requires detection of malicious activity by continuous monitoring of system critical resources and comparing it against the healthy behaviour. Once malicious activity is detected, generate an alert to initiate a mitigation strategy.

In the embedded security domain, there is a significant published literature on signature, anomaly and information flow-based detection methods~\cite{Wahab2017},~\cite{Coburn2005} as shown in Table~\ref{tab:security-requirements}. Within embedded architectures, these security mechanisms have been deployed at hardware and software layers managed by a \textit{Trusted Execution Environment} (TEE)~\cite{Sabt2015}.\vspace{0.10cm}


\noindent \textbf{\textcircled{4} RESPOND} to detected threats and malicious activities by planning and deploying an effective response and mitigation strategy to limit and reduce the impact of the cyber attack. \textit{Machine-to-Machine} (M2M) communication is an enabling technology for critical infrastructure~\cite{Wan2012}, which brought serious security challenges to secure, verify and avoid man-in-middle attacks in embedded systems. The existing embedded systems lack the capability to respond against attacks, making a need for active response against attacks a fundamental security requirement for cyber resilient embedded systems. Nevertheless, constantly evolving cyber attacks demand continuous re-evaluation for effective response and mitigation strategies.

Existing embedded security microarchitectures are largely focused on trust-based security and protection. They are limted and provide passive countermeasures such as watchdog timer, brownout reset, voltage and temperature monitoring and anti-tamper. Where, the vast majority system do not have any response mechanism and curtail such attacks using system reboot and reset. Nevertheless, trust management between device manufacturers and service providers is still a formidable challenge~\cite{Sharma2019}.
Clearly, there is a strong need for embedded response methods and microarchitectures that fulfil the security requirements of cyber resilient embedded system.\vspace{0.10cm}

\noindent \textbf{\textcircled{5} RECOVER} system data and resources back to the device healthy provisioned state, by repairing, updating and patching the system. However, effective cyber strategy requires identification of the causes and method of system failure by collecting evidence from the compromised system. It allow to establish, conduct and communicate critical administrative tasks with the actors involved, during the system life cycle, to effectively ensure and maintain safety and security of the critical systems.

The existing embedded security architectures are limited to the principles of reliability to achieve recovery, and thus are insufficient to provide a system-level information or evidence that can be used to improve the cyber strategy. They often make use of fault avoidance and fault tolerant design practises by incorporating redundant system resources and roll-back patches to return the system to a healthy state.

The mapping of existing embedded security approaches in Table~\ref{tab:security-requirements} clearly indicates a research gap and need for active response and recovery methods. Section~\ref{sec:defenselayer} extends this by discussing challenges and shortcomings of existing embedded security microarchitectures.


\section{Challenges and Shortcomings in existing Embedded Security Micorarchitectures}
\label{sec:defenselayer}
Embedded Security has been the subject of extensive research in the context of general-purpose computing, signal processing, programmable architectures and communications systems~\cite{Ravi2004}, with significant published work on various fine and coarse grained embedded security challenges~\cite{Ravi2004},~\cite{Apthorpe2017},~\cite{Serpanos2013},~\cite{Papp2015}. Security is often misapprehended by security architects and system designers as the addition of security features to make the system secure. Instead, security is a process that should be considered and managed throughout the life cycle of the embedded system specially for devices deployed in critical infrastructure. Therefore, this section first presents challenges of existing security microarchitectures:

\begin{itemize}
\item The majority of embedded security microarchitectures follow \textit{Device Trust Architecture}~\cite{Global2018}. It is a specification that provides a method to design and develop secure component technologies by building trust and secure services from the boot mechanism to the device operating system and application layer. Hence, the security of the system is strongly reliant on building and maintaining a strong \textit{chain of trust}~\cite{Ukil2011} which comprises of a series of nested assumptions and as vulnerable as its weakest link. If broken, compromises the security of the whole system. In the commercial domain, \textit{Secure Boot} is a well established and widely used secure component, which has been found vulnerable~\cite{McClintic2018},~\cite{Chen2017}.

\item A lack of clear ownership of device security, insufficient adoption of security-aware practises and an absence of baseline security requirements. Practically design engineers do not perceive themselves accountable for security requirements and effectively embedding them into the device life cycle. This includes a lack of formal security risk assessment, with management of security technology outsourced to third parties for design, development and formalised security patch management process. As a result, this integration of third party services leads to security inconsistencies and vulnerabilities.
\end{itemize}

These challenges have posed immense need for harnessing security, in compliance with cyber security regulations. This in turn, has compelled embedded security architects to design and deploy defence mechanisms that are often ad-hoc and passive in nature, targeting and mitigating certain attacks or classes of attacks after they have been discovered~\cite{Meng2018}. This approach may be effective to rectify software vulnerabilities or bugs through a software update, but insufficient to realise effective microarchitecture security which cannot be updated after release. The following are widely adopted embedded security methods has been found vulnerable due too poor usage of secure design practises, software vulnerabilities and microarchitectural weaknesses:

\begin{itemize}

\item \textbf{Trusted Computing}: Trusted software services uses cryptographic digital signatures to verify the integrity of the firmware and applications which has been exploited to gain access to the device~\cite{Chen2017}. This has occurred due to lack of roll-back prevention, as the system was using the same digital signature to verify the application. A similar attack has been performed against commercial TEE~\cite{ProjectZero2017}. 

\item \textbf{Processor virtualisation and logical isolation of resources}: In existing embedded security architectures, processor virtualisation has been used to achieve logical isolation between secure and non-secure system resources. This has been attacked through a covert cache-based attack, resulting in leaking of information using microarchitecture side-channels.
The recently demonstrated Spectre~\cite{Kiriansky2018} attack leverage speculative loads which circumvent access checks to read memory-resident secrets, transmitting them to an attacker using cache timing or other covert communication channels. Meltdown~\cite{Lipp2017} is another microarchitectural attack that exploits out-of-order execution to leak the target’s physical memory. These attacks exploit the fact that both secure and non-secure processes shares the same physical memory resource and pointer. Maene~\etal have proposed a data encryption mechanism~\cite{Maene2018} that allows automatic encryption and decryption of data between the main memory and cache though found infeasible due to large area overhead.
	
\item \textbf{Pointer Authentication}: To circumvent the microarchitecture side-channel leakage attacks, a pointer authentication mechanism has been introduced~\cite{Avanzi2017}. This guarantees the integrity of pointers by extending each pointer with authentication code, allowing verification using special instructions that are part of the code executing on the same physical computing resource and managed by the software. Similarly, to mitigate branch prediction attacks, deployment of separate stacks and their pointer registers has been introduced in ARM Cortex-M33 processors.
	
\item \textbf{Vulnerable system communication}: A security evaluation of the ARM TrustZone technology has demonstrated that it is possible to modify hardware security attributes and communication bus handshaking signals~\cite{Benhani2019}. This has demonstrated by integrating ARM TrustZone technology with reconfigurable hardware logic.
\end{itemize}

These microarchitectural weaknesses clearly indicates the need for cyber resilient embedded security microarchitecture that support active detection, response and recovery mechanisms to effectively realise diverse cyber security strategies through the life cycle of an embedded device. To this end, Section~\ref{sec:characteristics} proposes micro-architectural characteristics of a cyber resilient embedded system.
\section{Microarchitectural Characteristics of\\ a Cyber Resilient Embedded System}
\label{sec:characteristics}


As discussed, there is no active method in existing embedded microarchitectures to establish and maintain the security of a device once its trust is compromised. This leads to exposure of confidential data to the adversary, often without leaving any evidence trail. Considering the derived security requirements of cyber resilient embedded system (Table~\ref{tab:security-requirements}), security functionality is not limited to \textit{protection}. The device must \textit{detect} malicious cyber activities and attacks, \textit{respond} against them by deploying active countermeasures and \textit{recover} system back to its healthy state. These are crucial security requirements for embedded devices deployed in critical infrastructure as well as to facilitate forensic analysis, to study behaviour and method of cyber attacks.
Using existing embedded security microarchitectures, this is difficult and implausible to recover data due to lack of \textit{continuity of data stream}, \textit{runtime monitoring} and \textit{system-level visibility}. The following are proposed three core microarchitectural characteristics that shall allow to establish historical system data stream by continuous monitoring of system resources and activities, keeping track of events to achieve system-level visibility:

\begin{enumerate}

\item An \textbf{Independent Active Runtime System Security Manager} shall be responsible for \textit{protection}, \textit{detection}, \textit{response} and \textit{recovery} security functions while complimenting existing security mechanisms. It shall continuously monitor system resources, use gathered information to detect benign or malicious system behaviour, respond to detected malicious (system or resource-specific) activities by deploying active countermeasures and recover system back to its healthy state. It is crucial that \textit{system security manager} must be physically independent and isolated so its memory resources from the general purpose processor. This physical limiting of attack surface, will make the system robust and significantly less susceptible to software vulnerabilities and attacks as was in the case of the TEE. As TEE shares the same physical processor and memory resources with the general purpose processor. 
Effective realisation of this \textit{system security manager} requires resource-level visibility and monitoring of system's critical components which leads to the second characteristic.

\item An \textbf{Active Runtime Resource Monitors} shall actively monitor resource specific behaviour to \textit{detect} malicious activity and report it to the \textit{System Security Manager}. These active monitors are essential as embedded architectures are becoming complex, designers are consolidating diverse functionalities into a single application often involves mixing of sensitive data with non-sensitive data and physical actuation. These active monitors shall generate fine-grained resource specific information which would enable the \textit{system security manager} to articulate, analyse and evaluate system-level behaviours and initiate appropriate \textit{response} and \textit{recovery} strategies. In addition, this gathered information would facilitate continuity of data stream and to extract crucial information necessary to establish evidence of the caused security breach.

\item An \textbf{Active Response Manager} shall be responsible for implementing \textit{response} and \textit{recovery} embedded security requirements of a cyber resilient embedded system. It shall actively enforce and execute the response and recovery strategies initiated by the \textit{System Security Manager}. This involves initiating active countermeasures to mitigate and curtail the detected threat to maintain and ensure security of the system. Moreover, depending on the microarchitecture of the \textit{active runtime resource monitors}, the active response manager can enforce various system-level security strategies, where a compromised resource can be physically isolated from the system. This would allow opportunities to gracefully degrade the system functionality while maintaining critical services in next-generation critical infrastructure.

\end{enumerate}

A detailed System-on-Chip (SoC) platform architecture~\cite{SiddiquiIET2018},~\cite{Siddiqui2018SoCC} and security modelling approach~\cite{Hagan2018} that realises the proposed embedded microarchitectural characteristics of a cyber resilient embedded system have been published.
\section{Conclusion}
\label{sec:conclusion-future}

This paper has presented the increasing security challenges and requirements, in light of international cyber security regulations for intelligent connected technologies deployed in critical infrastructure. Embedded security requirements has been derived from these regulations to improve cyber resilience and achieve conformance. The paper establishes a strong need for embedded cyber resilience for smart technologies, due to lack of active detection, response and recovery security functionalities within existing embedded security systems. 

This is due to the majority of embedded security technologies being guided by trust, which has been compromised due to a lack of runtime monitoring and system-level visibility of resources and system activities. Therefore, this paper proposed three embedded microarchitectural characteristics, allowing independent active runtime system monitoring and active response functions to enhance, maintain and ensure secure operation during the life cycle of the device.

\bibliographystyle{IEEEtran}
\bibliography{IEEEabrv,IEEEReferences}
\end{document}